\documentclass[twocolumn,floatfix,showpacs,preprintnumbers,amsmath,amssymb,superscriptaddress,prl]{revtex4-1}
\usepackage[latin9]{inputenc}
\usepackage[active]{srcltx}
\usepackage{color}
\usepackage{float}
\usepackage{textcomp}
\usepackage{amsmath}
\usepackage{amssymb}
\usepackage{stackrel}
\usepackage{graphicx}
\usepackage{esint}

\makeatletter

\DeclareFontEncoding{LGR}{}{}
\DeclareRobustCommand{\greektext}{%
  \fontencoding{LGR}\selectfont\def\encodingdefault{LGR}}
\DeclareRobustCommand{\textgreek}[1]{\leavevmode{\greektext #1}}
\ProvideTextCommand{\~}{LGR}[1]{\char126#1}

\providecommand{\tabularnewline}{\\}

\usepackage{graphics}\usepackage{subfigure}\usepackage{longtable}\usepackage{pstricks}\usepackage{dcolumn}\usepackage{bm}

\makeatother

\begin{document}

\title{Strong magnetic frustration in Y$_{3}$Cu$_{9}$(OH)$_{19}$Cl$_{8}$:
a distorted kagome antiferromagnet}

\author{Pascal Puphal}

\affiliation{Physikalisches Institut, Goethe-University Frankfurt, 60438 Frankfurt
am Main, Germany}

\author{Michael Bolte}

\affiliation{Institut f\"ur Organische Chemie der Universit\"at Frankfurt, 60439
Frankfurt am Main, Germany}

\author{Denis Sheptyakov}

\affiliation{Laboratory for Neutron Scattering and Imaging, Paul Scherrer Institute,
5232 Villigen, Switzerland}

\author{Andrej Pustogow}

\affiliation{1. Physikalisches Institut, Stuttgart University, 70569 Stuttgart,
Germany}

\author{Kristin Kliemt}

\affiliation{Physikalisches Institut, Goethe-University Frankfurt, 60438 Frankfurt
am Main, Germany}

\author{Martin Dressel}

\affiliation{1. Physikalisches Institut, Stuttgart University, 70569 Stuttgart,
Germany}

\author{Michael Baenitz}

\affiliation{Max Planck Institute for Chemical Physics of Solids, D-01187 Dresden,
Germany}

\author{Cornelius Krellner}

\affiliation{Physikalisches Institut, Goethe-University Frankfurt, 60438 Frankfurt
am Main, Germany}
\begin{abstract}
\textcolor{black}{We present the crystal structure and magnetic properties
of }Y$_{3}$Cu$_{9}$(OH)$_{19}$Cl$_{8}$\textcolor{black}{, a stoichiometric
frustrated quantum spin system with slightly distorted kagome layers.
Single crystals of }Y$_{3}$Cu$_{9}$(OH)$_{19}$Cl$_{8}$\textcolor{black}{{}
were grown under hydrothermal conditions. The structure was determined
from single crystal X-ray diffraction and confirmed by neutron powder
diffraction. The observed structure reveals two different Cu-positions
leading to a slightly distored kagome layer in contrast to the closely
related YCu$_{3}$(OH)$_{6}$Cl$_{3}$. Curie-Weiss behavior at high-temperatures
with a Weiss-temperature $\theta_{W}$ of the order of $-100\,\text{K}$,
shows a large dominant antiferromagnetic coupling within the kagome
planes. Specific-heat and magnetization measurements on single crystals
reveal an antiferromagnetic transition at T$_{N}=2.2$~K indicating
a pronounced frustration parameter of $\theta_{W}/T_{N}\approx50$.
Optical transmission experiments on powder samples and single crystals
confirm the structural findings. Specific-heat measurements on YCu$_{3}$(OH)$_{6}$Cl$_{3}$
down to 0.4~K confirm the proposed quantum spin-liquid state of that
system. Therefore, the two Y-Cu-OH-Cl compounds present a unique setting
to investigate closely related structures with a spin-liquid state
and a strongly frustrated AFM ordered state, by slightly releasing
the frustration in a kagome lattice.}
\end{abstract}
\maketitle

\section{Introduction}

\textcolor{black}{Quantum spin systems with Cu$^{2+}$ ions are suitable
materials to study quantum many-body effects under variable conditions.
Prominent examples are low-dimensional materials with strong magnetic
frustrations. In these systems, a quantum spin-liquid state can be
realized at low temperatures which is a highly correlated state that
has no static magnetic order, despite sizeable magnetic interactions
\cite{Balens(2010)}. Compounds with decoupled antiferromagnetic kagome
layers are prototypical systems to search for an experimental realization
of the quantum spin-liquid state and Herbertsmithite, ZnCu$_{3}$(OH)$_{6}$Cl$_{2}$,
has become one of the most prominent materials in recent years \cite{Shores(2005),Mendels(2007),Han(2012),Helton(2007)}.}

\textcolor{black}{The dominant magnetic interaction in Herbertsmithite
is caused by Cu-O-Cu antiferromagnetic superexchange with an exchange
energy of J $\sim$17 meV, but no magnetic long-range order has been
observed down to T = 50~mK \cite{Helton(2007)}. Therefore, the spin-liquid
ground-state of this material could be investigated in great detail
(see e.g. a recent review in \cite{Norman(2016)}). One structural
drawback of Herbertsmithite is the intrinsic Zn-Cu-antisite disorder,
which makes it challenging to achieve a structurally perfect ZnCu$_{3}$(OH)$_{6}$Cl$_{2}$
crystal. Furthermore, the amount of antisite disorder is difficult
to quantify with X-ray scattering techniques \cite{Norman(2016),Freedman(2010)}.
Several structural variants including polymorphism with varying intersite
Cu-M mixing are common features of Cu-based kagome compounds such
as: Herbertsmithite - Kapellasite \cite{Helton(2007),Krause(2006)},
Mg-Herbertsmithite - Haydeaite \cite{Chu(2011)}, Volborthite - Vesignieit
\cite{Ishikawa(2012),Boldrin(2016)}, Francisite \cite{Prishchenko(2016)}
and Centennialite \cite{Sun(2016)-1}. From this point of view, novel
kagome systems with highly ordered crystal structures are essential
to uncover the intrinsic properties of the kagome antiferromagnet.
In addition, the frontier of Herbertsmithite is chemical doping \cite{Norman(2016)}
since Mazin et al. have proposed that a correlated Dirac metal can
be found in electron-doped Herbertsmithite which might be realized
by replacing Zn by a threevalent ion \cite{Mazin(2014)}. }

\textcolor{black}{Recently, W. Sun et al. reported on a non-hydrothermal
synthesis of YCu$_{3}$(OH)$_{6}$Cl$_{3}$ with LiOH and LiCl as
pH regulating additives \cite{Sun(2016)}, enclosed in an autoclave
to trap the crystal water. The reported structure is P$\bar{3}$m1
with one crystallographic copper position revealing a perfect two-dimensional
kagome lattice. The determined crystal structure is more reminiscent
to what was found in Kapellasite, a structural polymorph of Herbertsmithite
\cite{Colman(2008)}. They also show with Rietveld refinement of X-ray
diffraction (XRD) data and nuclear magnetic resonance (NMR) that there
is no significant Y - Cu exchange. In magnetic measurements on polycrystalline
samples with small amounts of impurity phases Sun et al. see no magnetic
order down to 2~K. The authors report that the absence of free water
in the starting mixtures has been proven to be a key factor in the
formation (and preservation) of this structural variant.}

\textcolor{black}{Here, we report on a different synthesis procedure
in the Y-Cu-OH-Cl system and we obtain single crystals of }Y$_{3}$Cu$_{9}$(OH)$_{19}$Cl$_{8}$\textcolor{black}{{}
from the hydrothermal method. For }Y$_{3}$Cu$_{9}$(OH)$_{19}$Cl$_{8}$\textcolor{black}{,
we find R$\bar{3}$ as the resulting structure with two distinct copper
positions and two fully occupied yttrium positions. As a consequence
}Y$_{3}$Cu$_{9}$(OH)$_{19}$Cl$_{8}$\textcolor{black}{{} presents
the stoichiometric case of a slightly distorted kagome system, leading
to the stabilization of magnetic order at T$_{N}=2.2\,$K but a large
portion of the spin degrees of freedom remain fluctuating. Therefore,
}Y$_{3}$Cu$_{9}$(OH)$_{19}$Cl$_{8}$\textcolor{black}{{} and YCu$_{3}$(OH)$_{6}$Cl$_{3}$
are ideal systems to investigate the change of a spin-liquid state
to a strongly frustrated AFM ordered state, by slightly releasing
the frustration in a kagome lattice.}

\section{Experimental}

\subsection{Synthesis}

\textcolor{black}{Single crystals of }Y$_{3}$Cu$_{9}$(OH)$_{19}$Cl$_{8}$\textcolor{black}{{}
were prepared in a hydrothermal Parr 4625 autoclave with a 575 ml
filling capacity operated by a Parr 4842 power supply including a
982 Watlow controller. The crystals have blue to green colour and
a hexagonal shape, typical sizes are up to 1 x 1 x 0.25 mm$^{3}$.
For the crystallization, we placed duran glass ampoules filled with
the solution in the autoclave and filled it with distilled water to
ensure the same pressure as in the ampoules. The ampoules were loaded
with 0.59~g Y$_{2}$O$_{3}$, 0.82~g CuO, 0.89~g CuCl$_{2}\cdot2$
(H$_{2}$O) and 10~ml distilled water and then sealed at air. The
autoclave was heated up to 270\textdegree C in four hours and subsequently
cooled down to 260\textdegree C with 0.05 K/h, followed by a fast
cooling to room temperature. Afterwards, the ampoules were opened
and the content was filtered with distilled water. Unlike in the synthesis
of the similar compounds (Herbertsmithite, MgCu$_{3}$(OH)$_{6}$Cl$_{2}$
Mg-Herbertsmithite \cite{Colman(2011)}, and CdCu$_{3}$(OH)$_{6}$Cl$_{2}$
\cite{MCQueen(2011)}) there is no need to use excess Y in the growth,
since the Y$_{2}$O$_{3}$ can easily be solved in chloridic solutions
\cite{Meyer (1989)}. Attempts with an excess of YCl$_{3}$ in the
solution in fact lead to a formation of Y(OH)$_{3}$. In addition,
only the $x=1$ stoichiometry of Y$_{x}$Cu$_{4-x}$(OH)$_{6.33}$Cl$_{2.77}$
forms and we have not observed any phase with a $x$-value below 1,
in contrast to the other compounds of this family.}

\textcolor{black}{Furthermore to compare the magnetic ground state
of the two structural variants, we have reproduced the synthesis of
polycrystalline YCu$_{3}$(OH)$_{6}$Cl$_{3}$ with the flux method
from Ref. \cite{Sun(2016)}, where Y(NO$_{3}$)$_{3}\cdot6$H$_{2}$O
melts at 50\textdegree C, starting to form complexes \cite{Melnikov(2013)}.
We analysed the obtained powder with laser ablation - inductively
coupled plasma - mass spectrometry (LA-ICP-MS). We found only a few
ppm of Lithium and could therefore exclude lithium incorporation in
YCu$_{3}$(OH)$_{6}$Cl$_{3}$.}

\subsection{Characterization}

\textcolor{black}{For the single crystal structure determination the
data were collected at 173~K on a STOE IPDS II two-circle diffractometer
with a Genix Microfocus tube with mirror optics using Mo K$_{\alpha}$
radiation ($\lambda=0.71073\textrm{\AA}$). The data were scaled using
the frame scaling procedure in the X-AREA program system \cite{Stoe(2002)}.
The structure was solved by direct methods using the program SHELXS
\cite{Sheldrick(2008)} and refined against $F^{2}$ with full-matrix
least-squares techniques using the program SHELXL-97 \cite{Sheldrick(2008)}.
The H atoms bonded to O2, O3, and O4 were found in a difference map
and were isotropically refined with the O-H distance restrained to
0.84(1)Å, whereas the H1 atom bonded to O1 was geometrically positioned
and refined using a riding model. The crystal was twinned about (-1
0 0/1 1 0/0 0 -1) with a fractional contribution of 0.601(4) for the
major domain. The space group R$\bar{3}$ was chosen, because the
structure proved to be centrosymmetric (thus R3 could be excluded)
and no hints for any mirror planes were detected (excluding space
groups R3m and R$\bar{3}$m). We measured neutron powder diffraction
at the high-resolution powder neutron diffractometer HRPT \cite{Fischer(2000)},
at the Paul Scherrer Institute in Villigen. An amount of $\sim1$
g of }Y$_{3}$Cu$_{9}$(OH)$_{19}$Cl$_{8}$\textcolor{black}{{} was
enclosed into a vanadium can with an inner diameter of 6 mm and the
measurement was carried out at room temperature}\textcolor{black}{\small{}
with a }\textcolor{black}{wavelength of $\lambda=1.494\textrm{\AA}$.
The Rietveld refinement, of the neutron data, accomplishing the crystal
structure was done using the fullprof suite \cite{Rodriguez(1993)}.
Low-temperature diffraction data were collected with a Siemens D-500
diffractometer with Cu K$_{\alpha}$ radiation ($\lambda=1.5406\,\textrm{Å}$),
here, the sample was placed on a Cu-sample holder in a Lakeshore M-22
closed cycle refridgerator. }

\textcolor{black}{For the optical characterization a Bruker Fourier-transform
infrared spectrometer and a Woollam spectroscopic ellipsometer have
been utilized. Optical transmission experiments in the mid-infrared
range were performed with KBr powder pellets whereas thin flakes were
used in the visible/UV range. The discussed features were also observed
in single-crystalline samples of different thickness ($d=15-70$ \textmu m)
proving them as bulk properties. }

\textcolor{black}{The specific-heat data and magnetic measurements
were collected with the standard options of a Physical Property Measurement
System from Quantum Design in a temperature range of 0.4 to 300~K.}

\section{results and discussion}

\subsection{Crystal structure}

\begin{table}[H]
\caption{{\footnotesize{}\label{structural parameters}Crystal structure parameters
of }Y$_{3}$Cu$_{9}$(OH)$_{19}$Cl$_{8}${\footnotesize{} refined
from X-ray single crystal diffraction data measured at 173 K. Space
group R$\bar{3}$ (\# 148). The unit cell parameters are $a=b=11.5350(8)\textrm{\,Å}$
and $c=17.2148(12)\textrm{\,Å}$. All positions are fully occupied
except H1, which has an occupancy of 1/6.}}

\begin{ruledtabular}

{\footnotesize{}}%
\begin{tabular}{llllll}
 & Wyck. & {\small{}x/a} & {\small{}y/b} & {\small{}z/c} & {\small{}U {[}$\textrm{Å}^{2}${]}}\tabularnewline
{\footnotesize{}Cu1} & 18f & {\footnotesize{}0.66311(6)} & {\footnotesize{}0.82526(6)} & {\footnotesize{}0.50349(3)} & {\footnotesize{}0.00721(19)}\tabularnewline
{\footnotesize{}Cu2} & 9d & {\footnotesize{}0.5} & {\footnotesize{}1} & {\footnotesize{}0.5} & {\footnotesize{}0.0072(2)}\tabularnewline
{\footnotesize{}Y1} & 6c & {\footnotesize{}0.3333} & {\footnotesize{}0.6667} & {\footnotesize{}0.53850(4)} & {\footnotesize{}0.0075(2)}\tabularnewline
{\footnotesize{}Y2} & 3b & {\footnotesize{}1} & {\footnotesize{}1} & {\footnotesize{}0.5} & {\footnotesize{}0.0092(3)}\tabularnewline
{\footnotesize{}Cl1} & 18f & {\footnotesize{}0.66466(12)} & {\footnotesize{}1.00242(12)} & {\footnotesize{}0.61719(6)} & {\footnotesize{}0.0132(3)}\tabularnewline
{\footnotesize{}Cl2} & 6c & {\footnotesize{}1} & {\footnotesize{}1} & {\footnotesize{}0.33854(11)} & {\footnotesize{}0.0121(4)}\tabularnewline
{\footnotesize{}O1} & 3a & {\footnotesize{}0.3333} & {\footnotesize{}0.6667} & {\footnotesize{}0.6667} & {\footnotesize{}0.042(3)}\tabularnewline
{\footnotesize{}H1} & 18f & {\footnotesize{}0.404} & {\footnotesize{}0.7382} & {\footnotesize{}0.6561} & {\footnotesize{}0.063}\tabularnewline
{\footnotesize{}O2} & 18f & {\footnotesize{}0.8113(4)} & {\footnotesize{}0.8026(4)} & {\footnotesize{}0.54394(18)} & {\footnotesize{}0.0079(7)}\tabularnewline
{\footnotesize{}H2} & 18f & {\footnotesize{}0.813(6)} & {\footnotesize{}0.801(7)} & {\footnotesize{}0.5927(7)} & {\footnotesize{}0.012}\tabularnewline
{\footnotesize{}O3} & 18f & {\footnotesize{}0.5308(4)} & {\footnotesize{}0.6623(5)} & {\footnotesize{}0.55719(18)} & {\footnotesize{}0.0086(6)}\tabularnewline
{\footnotesize{}H3} & 18f & {\footnotesize{}0.555(6)} & {\footnotesize{}0.654(7)} & {\footnotesize{}0.6018(15)} & {\footnotesize{}0.013}\tabularnewline
{\footnotesize{}O4} & 18f & {\footnotesize{}0.5089(4)} & {\footnotesize{}0.8403(4)} & {\footnotesize{}0.46536(18)} & {\footnotesize{}0.0070(6)}\tabularnewline
{\footnotesize{}H4} & 18f & {\footnotesize{}0.492(6)} & {\footnotesize{}0.818(6)} & {\footnotesize{}0.4185(11)} & {\footnotesize{}0.011}\tabularnewline
\end{tabular}{\footnotesize \par}

\end{ruledtabular} 
\end{table}

\begin{figure}[H]
\includegraphics[width=1\columnwidth]{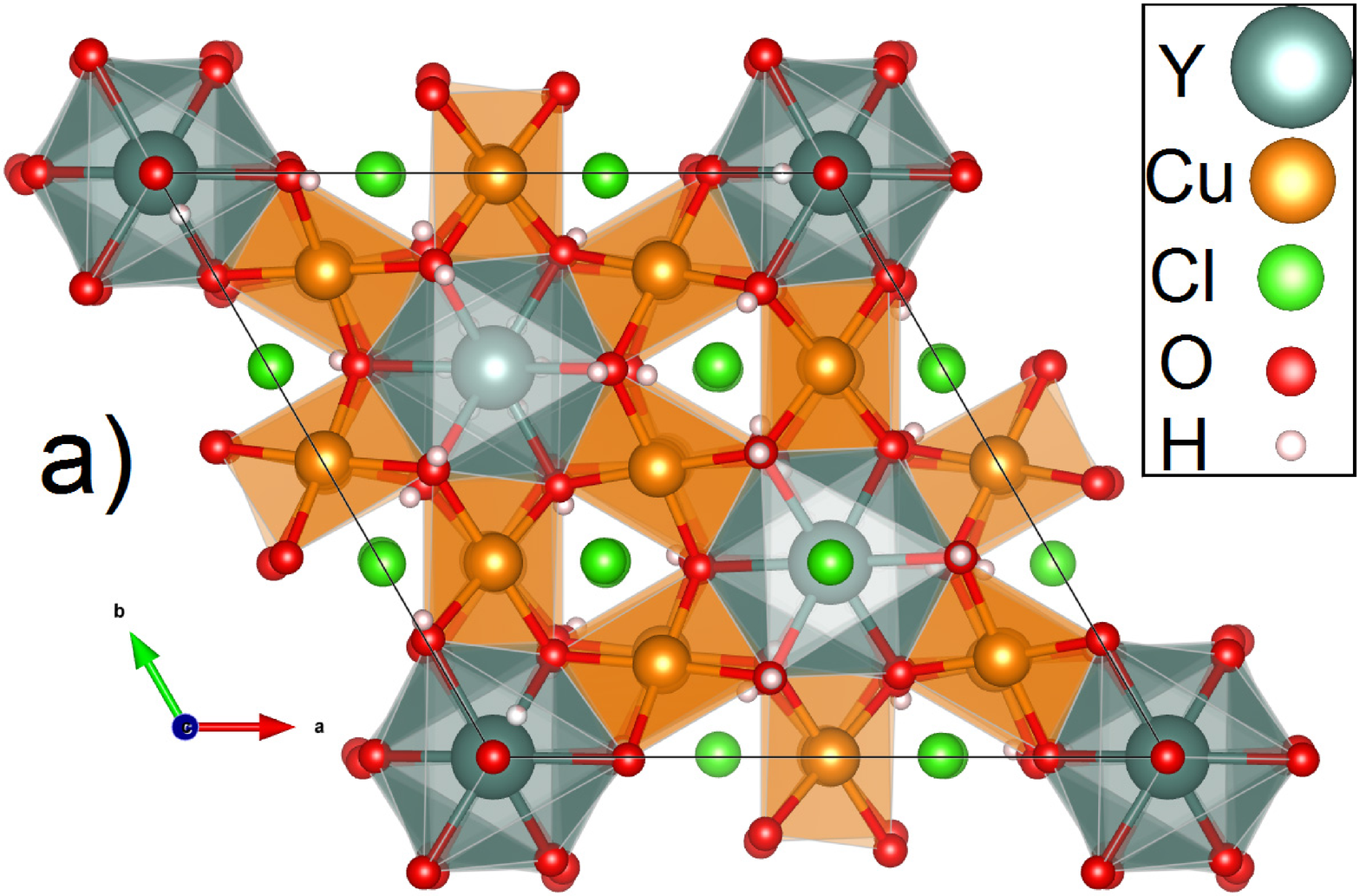}

\includegraphics[width=1\columnwidth]{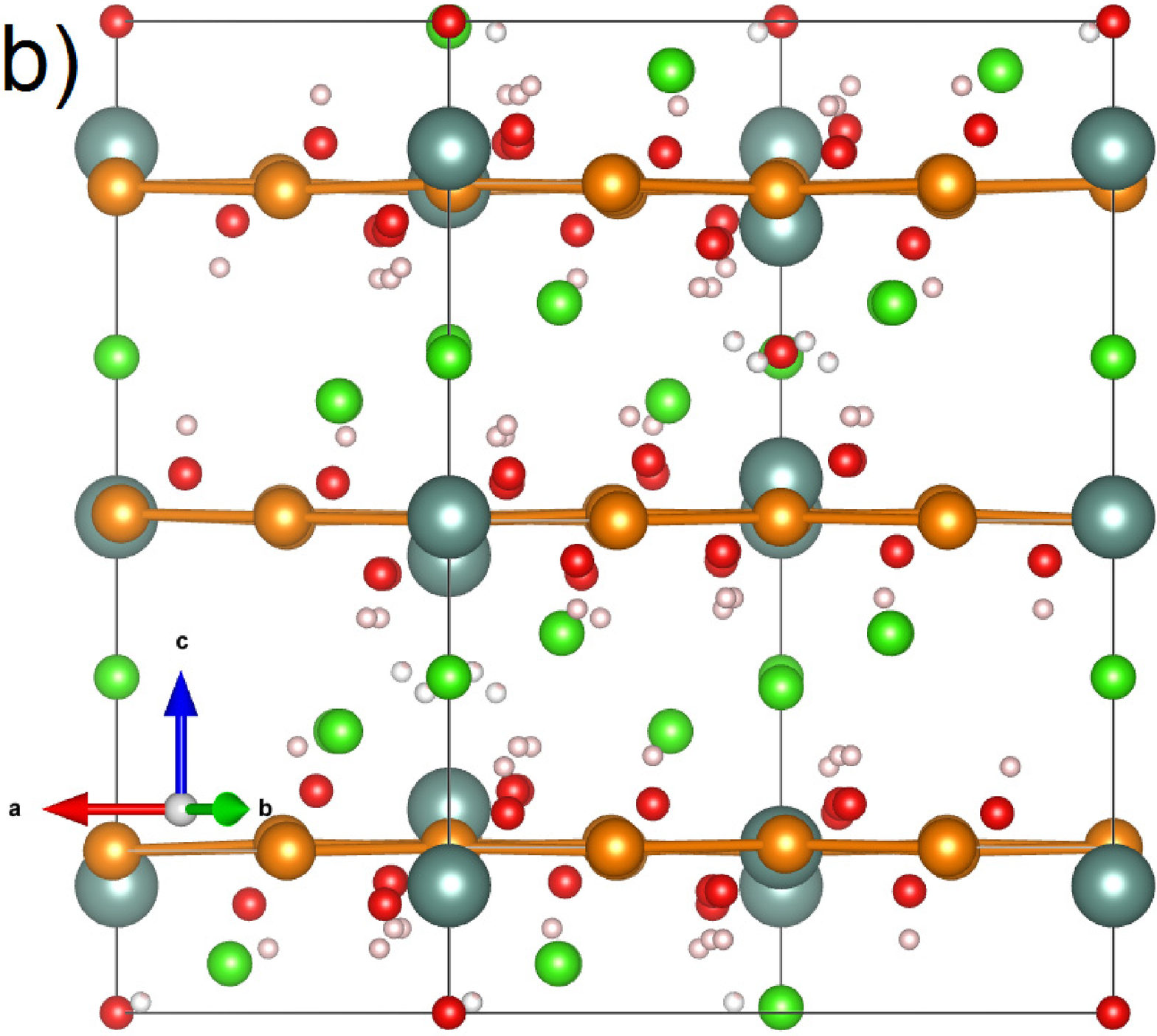}

\includegraphics[width=1\columnwidth]{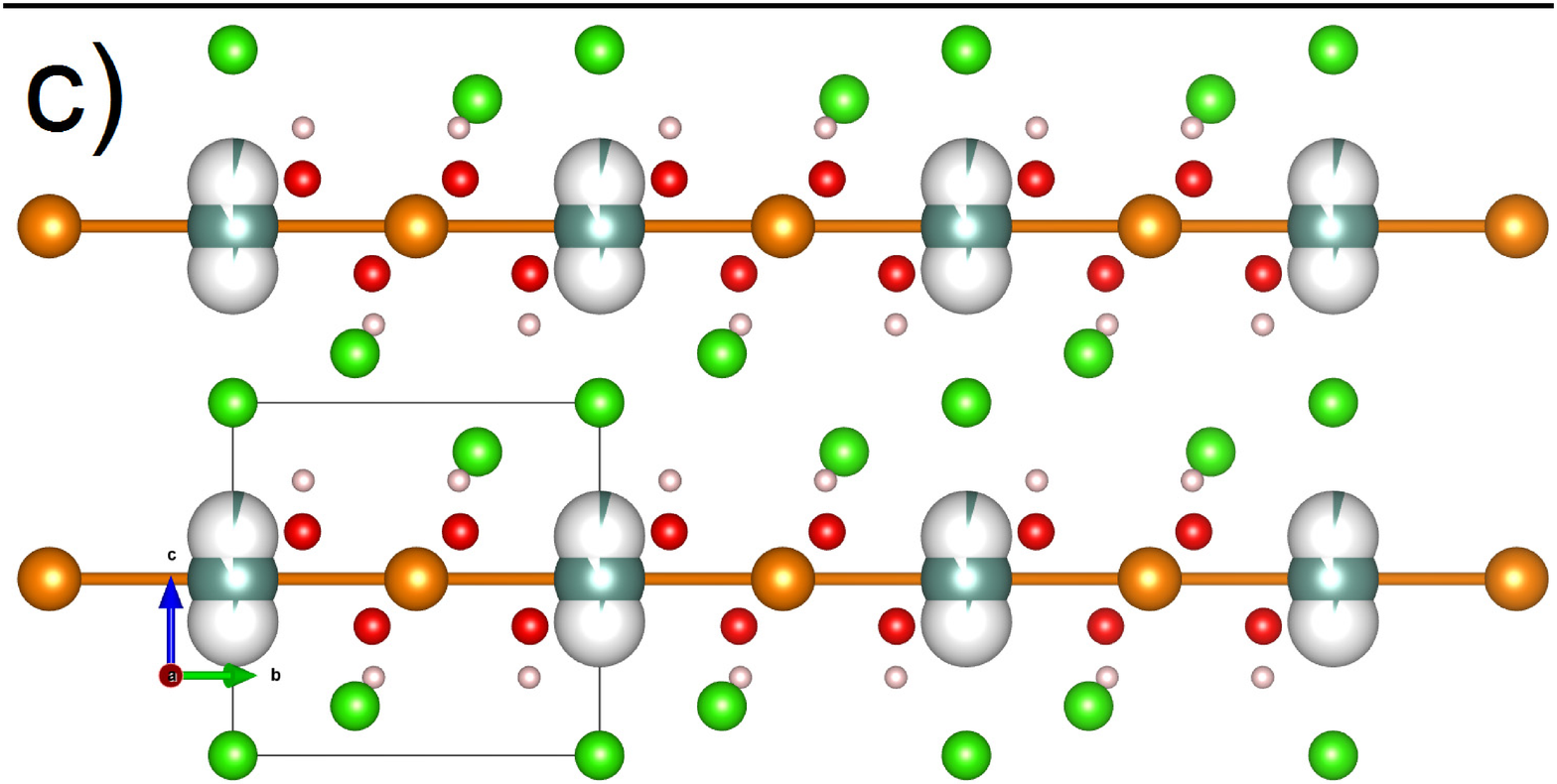}

\caption{\textcolor{black}{\footnotesize{}\label{structure}Structure model
of }{\footnotesize{}Y$_{3}$Cu$_{9}$(OH)$_{19}$Cl$_{8}$(a,b)}\textcolor{black}{\footnotesize{}
in comparison to YCu$_{3}$(OH)$_{6}$Cl$_{3}$ (c). a) Top view of
the kagome layer in }{\footnotesize{}Y$_{3}$Cu$_{9}$(OH)$_{19}$Cl$_{8}$}\textcolor{black}{\footnotesize{}
crystallizing in the R$\bar{3}$ structure. b) Side view of the kagome
layer in }{\footnotesize{}Y$_{3}$Cu$_{9}$(OH)$_{19}$Cl$_{8}$}\textcolor{black}{\footnotesize{}
stacked along the c-direction, where the slight buckling of the Cu-atoms
due to two crystallographic positions is apparent. c) Side view of
the kagome layer in YCu$_{3}$(OH)$_{6}$Cl$_{3}$ with the P$\bar{3}$m1
structure type \cite{Sun(2016)} with a partially filled second position
of Y.}}
\end{figure}

\textcolor{black}{The crystal structure of }Y$_{3}$Cu$_{9}$(OH)$_{19}$Cl$_{8}$\textcolor{black}{{}
is different from the theoretically proposed structure for electron-doped
Herbertsmithite \cite{Mazin(2014)} as well as the structure reported
by \cite{Sun(2016)}. The additional electron from yttrium is bound
with additional Cl$^{-}$/ OH$^{-}$ anions, so electron doping of
Herbertsmithite was not successful in this structure. In agreement
with Ref. \cite{Guterding(2016)}, the Y-atoms are not incorporated
into the interlayer site but in the kagome layer. Thus, }Y$_{3}$Cu$_{9}$(OH)$_{19}$Cl$_{8}$\textcolor{black}{{}
consists of repeating kagome layers. The distances between the Cu
atoms are 3.2498 $\textrm{\AA}$/3.3683 $\textrm{\AA}$/3.3762 $\textrm{\AA}$
within the layer and 5.8607 $\textrm{\AA}$/ 5.6788$\textrm{\AA}$
between the layers. In comparison, Herbertsmithite has Cu distances
of 3.416 $\textrm{\AA}$ and 5.087 $\textrm{\AA}$ \cite{Freedman(2010)}.
In Figure \ref{structure}, the obtained crystal structure of }Y$_{3}$Cu$_{9}$(OH)$_{19}$Cl$_{8}$\textcolor{black}{{}
is shown.}\textcolor{red}{{} }\textcolor{black}{A comparison of the
two similar structures can be made from Figure \ref{structure} b)
and c). In contrast to }Y$_{3}$Cu$_{9}$(OH)$_{19}$Cl$_{8}$\textcolor{black}{$_{3}$,
the P$\bar{3}$m1 structure has a partial disorder, for the two inequivalent
Y positions, displayed as white parts of the sphere. The full occupancy
of these two crystallographic Y positions in }Y$_{3}$Cu$_{9}$(OH)$_{19}$Cl$_{8}$\textcolor{black}{{}
causes the Cu-atoms to be slightly misaligned from a perfect plane.
W. Sun et al. observed an increasing occupation of these disordered
yttrium-atoms with decreasing temperature showing a tendency to the
}R$\bar{3}$\textcolor{black}{{} structure when lowering the temperature\cite{Sun(2016)}.}

\begin{figure}[h]
\includegraphics[width=1\columnwidth]{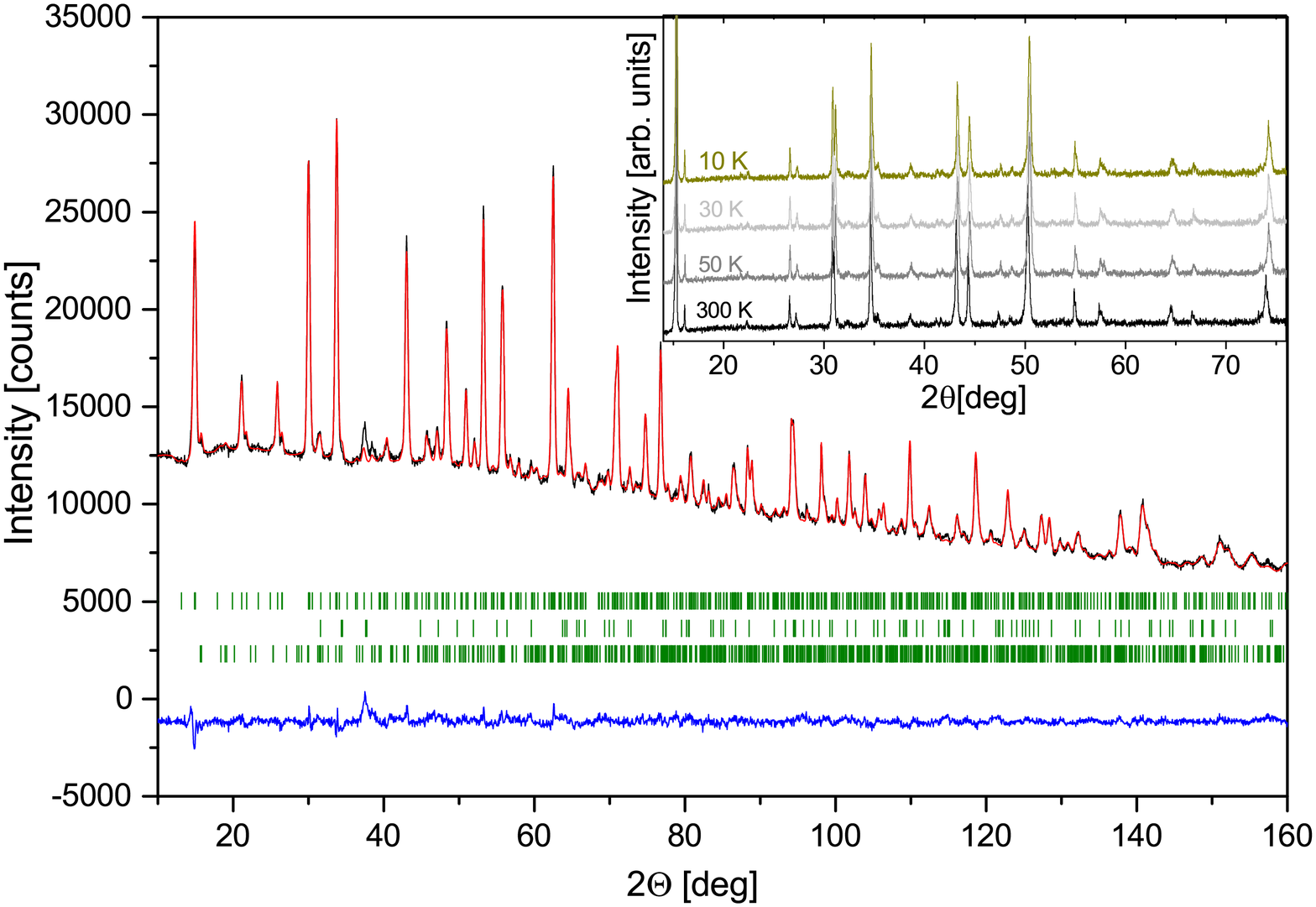}

\caption{{\footnotesize{}\label{neutron}Rietveld refinement of the crysta}\textcolor{black}{\footnotesize{}l
structure parameters from neutron powder diffraction data measured
at $T=295$\,K. The observed intensity (black), calculated profile
(red), and difference curve (blue) are shown of a powder sample. The
rows of ticks at the bottom correspond to the calculated diffraction
peak positions of the phases (from top to bottom): }{\footnotesize{}Y$_{3}$Cu$_{9}$(OH)$_{19}$Cl$_{8}$}\textcolor{black}{\footnotesize{},
CuO 2.5(2) wt\%. and Cu$_{2}$(OH)$_{3}$Cl 5.8(2) wt\%. Inset: Low-temperature
powder X-ray diffraction at different temperatures. }}
\end{figure}

A \textcolor{black}{refinement plot based on the neutron powder diffraction
data, at room temperature taken on 1~g of }Y$_{3}$Cu$_{9}$(OH)$_{19}$Cl$_{8}$\textcolor{black}{{}
is shown in Figure \ref{neutron}. The structure concept was obtained
by the above described single crystal X-ray data and the refinement
of the neutron data confirmed that structure solution. A small amount
of unreacted CuO and Cu$_{2}$(OH)$_{3}$Cl was observed in the large
powder sample which formed due to the off-stoichiometric synthesis
conditions. The unaccounted reflex part at 37\textdegree{} can also
be accounted to some impurity phase.}

\textcolor{black}{It should be noted, that neutron scattering lengths
of Y and Cu are in fact very close to each other: 7.75 and 7.718~fm,
which makes them practically indistinguishable in neutron refinements.
We therefore concluded the stoichiometric ordering without sizeable
Y-Cu site exchange from the single-crystal X-ray refinement. There,
the scattering cross sections for Y ($Z=39$) and Cu ($Z=29$) are
sufficiently different.}

\begin{table}[H]
\caption{{\footnotesize{}\label{structural parameters-1}Crystal structure
parameters of }Y$_{3}$Cu$_{9}$(OH)$_{19}$Cl$_{8}${\footnotesize{}
refined from neutron diffraction data measured at 295~K. Space group
R$\bar{3}$ (\# 148). The unit cell parameters are $a=b=11.5528(3)\textrm{\,Å}$
and $c=17.2216(5)\textrm{\,Å}$. All positions are fully occupied
except H1, which has an occupancy of 1/6.}}

\begin{ruledtabular}

{\footnotesize{}}%
\begin{tabular}{llllll}
 & {\small{}Wyck.} & {\small{}x/a} & {\small{}y/b} & {\small{}z/c} & {\small{}U {[}$\textrm{Å}^{2}${]}}\tabularnewline
{\small{}Cu1} & {\small{}18f} & {\small{}0.6662(6)} & {\small{}0.8286(7)} & {\small{}0.5036(3)} & {\small{}0.0121(4)}\tabularnewline
{\small{}Cu2} & {\small{}9d} & {\small{}0.5} & {\small{}1} & {\small{}0.5} & {\small{}0.0121(4)}\tabularnewline
{\small{}Y1} & {\small{}6c} & {\small{}0.3333} & {\small{}0.6667} & {\small{}0.5393(3)} & {\small{}0.0074(9)}\tabularnewline
{\small{}Y2} & {\small{}3b} & {\small{}1} & {\small{}1} & {\small{}0.5} & {\small{}0.0074(9)}\tabularnewline
{\small{}Cl1} & {\small{}18f} & {\small{}0.6636(7)} & {\small{}0.9959(7)} & {\small{}0.61849(17)} & {\small{}0.0220(5)}\tabularnewline
{\small{}Cl2} & {\small{}6c} & {\small{}1} & {\small{}1} & {\small{}0.3379(5)} & {\small{}0.0220(5)}\tabularnewline
{\small{}O1} & {\small{}3a} & {\small{}0.33333} & {\small{}0.66667} & {\small{}0.66667} & {\small{}0.022(3)}\tabularnewline
{\small{}H1} & {\small{}18f} & {\small{}0.404} & {\small{}0.7382} & {\small{}0.66667} & {\small{}0.0296(15)}\tabularnewline
{\small{}O2} & {\small{}18f} & {\small{}0.8092(8)} & {\small{}0.8023(8)} & {\small{}0.5435(3)} & {\small{}0.0119(4)}\tabularnewline
{\small{}H2} & {\small{}18f} & {\small{}0.7874(15)} & {\small{}0.8040(13)} & {\small{}0.5946(8)} & {\small{}0.0296(15)}\tabularnewline
{\small{}O3} & {\small{}18f} & {\small{}0.5295(8)} & {\small{}0.6618(12)} & {\small{}0.5578(3)} & {\small{}0.0119(4)}\tabularnewline
{\small{}H3} & {\small{}18f} & {\small{}0.5625(16)} & {\small{}0.668(2)} & {\small{}0.6099(6)} & {\small{}0.0296(15)}\tabularnewline
{\small{}O4} & {\small{}18f} & {\small{}0.5091(8)} & {\small{}0.8421(8)} & {\small{}0.4642(3)} & {\small{}0.0119(4)}\tabularnewline
{\small{}H4} & {\small{}18f} & {\small{}0.5024(17)} & {\small{}0.8255(16)} & {\small{}0.4076(7)} & {\small{}0.0296(15)}\tabularnewline
\end{tabular}{\footnotesize \par}

\end{ruledtabular} 
\end{table}

\textcolor{black}{We have used X-ray diffraction to prove the ideal
cation order of the Y and Cu, and neutron diffraction to identify
and precisely refine the positions of the hydrogen atoms in the structure,
since the scattering contrast of hydrogen is sufficiently high in
neutron diffraction (scattering length is negative, $b_{H}=-$3.739~fm,
as opposed to +7.75, +7.718, +5.803, and +9.577 fm for Y, Cu, O, and
Cl, correspondingly), which allowed for a refinement of the atomic
positions of hydrogen in the structure. Using a sample produced with
chemicals containing natural hydrogen (and not deuterium, as usually
done for a neutron diffraction study) did of course condition a rather
high background of incoherent scattering in the pattern, thus a longer
acquisition time was needed to achieve sufficient statistics for a
reliable refinement. We note that using a third Cl place instead of
the proposed O1 and H1 places would lead to the stoichiometry YCu$_{3}$(OH)$_{6}$Cl$_{3}$
similar to the structure of Ref. \cite{Sun(2016)}. Neutron refinement
indicates the absence of the H1 atom, thus only fully occupied O1,
the result was not taken into account due to charge balance.}

The \textcolor{black}{phase stability at various temperatures was
investigated using powder X-ray diffraction (PXRD) data at 10 K -
300 K in 20 K steps. The inset in Fig. \ref{neutron} reveals that
no structural phase transition could be resolved down to 10 K.}

\begin{figure}[h]
\includegraphics[width=1\columnwidth]{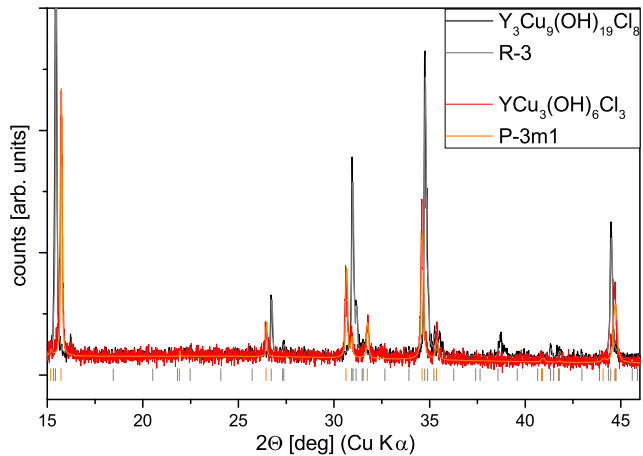}

\caption{\textcolor{black}{\footnotesize{}\label{diff comp}Powder X-ray diffraction
data of the hydrothermally grown }{\footnotesize{}Y$_{3}$Cu$_{9}$(OH)$_{19}$Cl$_{8}$}\textcolor{black}{\footnotesize{}
(black) with an underlying refinement of the R$\bar{3}$ structure
(grey) compared to the diffraction of a YCu$_{3}$(OH)$_{6}$Cl$_{3}$
sample (red) with the P$\bar{3}$m1 structure refinement (orange).}}
\end{figure}

\textcolor{black}{The two structure types are further compared in
Fig. \ref{diff comp}, where we present PXRD data of the two Y-Cu-OH-Cl
compounds. For that we have reproduced the synthesis method of Ref.
\cite{Sun(2016)}. From Fig. \ref{diff comp} it is obvious, that
a discrimination between the two compounds can be easily done using
PXRD.}

\subsection{Optical measurements}

\begin{figure}[H]
\includegraphics[width=1\columnwidth]{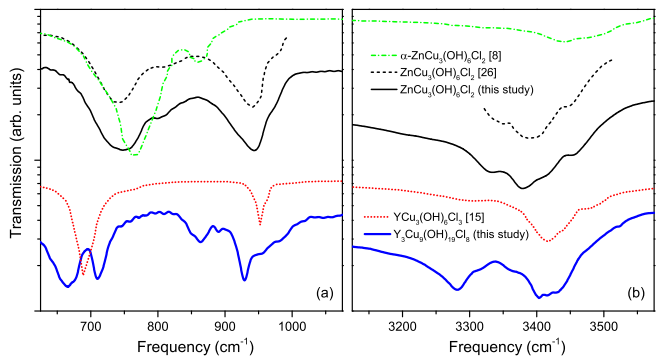}

\caption{\textcolor{black}{\footnotesize{}\label{trans}Powder transmission
measurements in the mid-infrared spectral range reveal pronounced
differences between the Kagome-lattice compounds YCu$_{3}$(OH)$_{6}$Cl$_{3}$
(red dotted line), }{\footnotesize{}Y$_{3}$Cu$_{9}$(OH)$_{19}$Cl$_{8}$}\textcolor{black}{\footnotesize{}
(blue) , \textgreek{a}-ZnCu$_{3}$(OH)$_{6}$Cl$_{2}$ (Kapellasite,
green) and ZnCu$_{3}$(OH)$_{6}$Cl$_{2}$ (Herbertsmithite, black).
For the latter compound, our data (full line) match perfectly with
literature (dashed line) \cite{Braithwaite(2004)}. Similar as for
the two ZnCu$_{3}$(OH)$_{6}$Cl$_{2}$ polymorphs, there are striking
differences in the optical response between YCu$_{3}$(OH)$_{6}$Cl$_{3}$
from Ref. \cite{Sun(2016)} and} {\footnotesize{}Y$_{3}$Cu$_{9}$(OH)$_{19}$Cl$_{8}$}\textcolor{black}{\footnotesize{}
which reflects the structural differences like bond lengths and angles.
While the vibrational features in the range 700 \textendash{} 1000
cm$^{-1}$ are associated with CuO-H deformations (a) and thus give
information about the kagome layer, O-H stretching vibrations are
observed between 3200 and 3500 cm-1 (b) \cite{Braithwaite(2004),Sushkov(2016)}. }}
\end{figure}

\textcolor{black}{As these materials are insulators, the electrodynamic
response in the mid-infrared range is dominated by phonons. Following
the assignment of previous studies, the features around 700 - 1000
and 3200 - 3500 cm$^{-1}$ are related to CuO-H deformations in the
Kagome layer and O-H stretching vibrations, respectively \cite{Braithwaite(2004),Sushkov(2016)}.
In Fig. \ref{trans}, we plot the optical transmission spectra for
the related Kagome-lattice compounds YCu$_{3}$(OH)$_{6}$Cl$_{3}$
(from Ref. \cite{Sun(2016)}), }Y$_{3}$Cu$_{9}$(OH)$_{19}$Cl$_{8}$\textcolor{black}{,
\textgreek{a}-ZnCu$_{3}$(OH)$_{6}$Cl$_{2}$ (Kapellasite, Ref. \cite{Krause(2006)})
and ZnCu$_{3}$(OH)$_{6}$Cl$_{2}$ (Herbertsmithite, our study and
Ref. \cite{Braithwaite(2004)}). There are striking differences between
the two ZnCu$_{3}$(OH)$_{6}$Cl$_{2}$ polymorphs, such as the absence
(or strong softening) of the 940 cm$^{-1}$ mode in Kapellasite, which
reflects the structural differences like bond lengths and angles.
Similarly, we clearly observe severe discrepancies between YCu$_{3}$(OH)$_{6}$Cl$_{3}$
from Ref. \cite{Sun(2016)} and }Y$_{3}$Cu$_{9}$(OH)$_{19}$Cl$_{8}$\textcolor{black}{:
Comparing the two we find a shift and splitting of the modes around
928 cm$^{-1}$ and 700 cm$^{-1}$, respectively, as well as additional
peaks at 860 cm$^{-1}$ and 3280 cm$^{-1}$ , confirming the structural
differences in these compounds, which is consistent with the discussed
structure details.}

\begin{figure}[H]
\includegraphics[width=1\columnwidth]{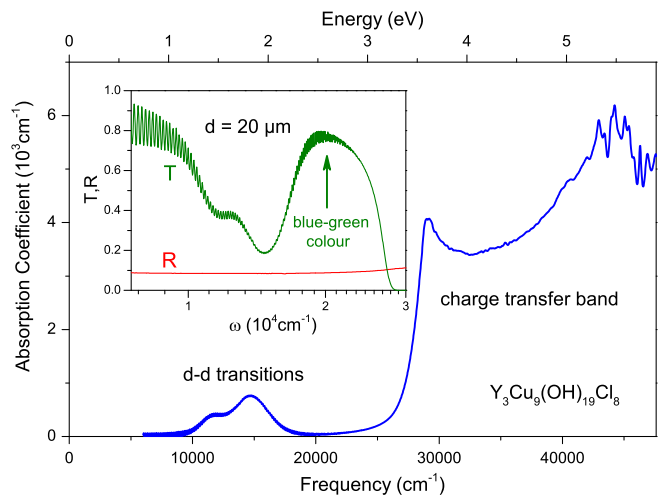}

\caption{{\footnotesize{}\label{UV}The VIS-UV absorption spectrum of Y$_{3}$Cu$_{9}$(OH)$_{19}$Cl$_{8}$
reveals the typical features of transition-metal oxides with d-d transitions
in the visible range and the charge-transfer band in the UV. Inset:
Pronounced Fabry-Perot fringes occur in the transmission data yielding
a refractive index of $n\approx1.7-1.8$ in agreement with the optical
reflectivity that is basically flat and featureless in the visible
range. As a consequence of the absorption in the long-wavelength range,
the largest transmission occurs at around 20,000 cm$^{-1}$ causing
the blue-green colour of the crystals. }}
\end{figure}

In addition, we have performed optical transmission experiments on
thin single crystals to study the electronic excitations in the visible
and ultraviolet spectral ranges. The absorption spectrum shown in
Fig. \ref{UV} reveals the crystal field splitting of Cu $d$-orbitals
due to Jahn-Teller distortions in the visible and the charge-transfer
band with a van-Hove like peak at 29~100 cm$^{-1}$ (3.6 eV) in the
UV. The d-d transitions have maxima at 11~500 and 14~700 cm$^{-1}$
which is in the typical range of copper-oxides \cite{Hwu(2002),Sala(2011)}.
The blue-green colour of the crystals stems from the transmission
maximum at 20~000 cm$^{-1}$. From our optical experiments at lower
frequencies we found no indication of the Dirac bands \cite{Mazin(2014)}.
This agrees with the stoichiometry as the excess charge of the Y$^{3+}$
cations is balanced by Cl$^{-}$/OH$^{-}$ anions and the kagome layer
is actually not doped. 

\subsection{Magnetic susceptibility and specific heat}

\begin{figure}[H]
\includegraphics[width=1\columnwidth]{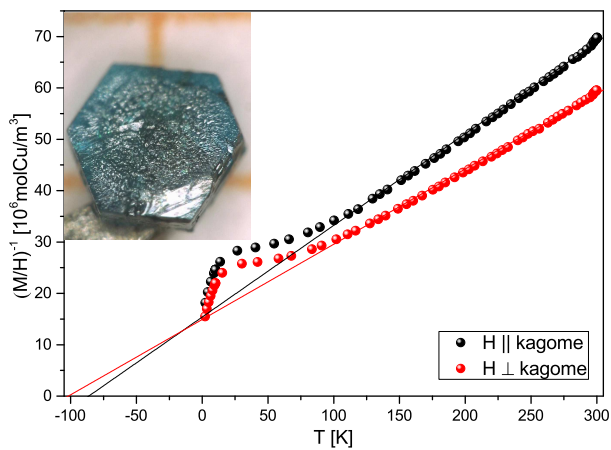}

\caption{\textcolor{black}{\footnotesize{}\label{M-1}Inverse magnetization
of a }{\footnotesize{}Y$_{3}$Cu$_{9}$(OH)$_{19}$Cl$_{8}$}\textcolor{black}{\footnotesize{}
single crystal ($m=1.34$ mg) in SI units for the two field-directions,
parallel and perpendicular to the kagome layers at $H=1$~T from
300 to 1.8~K. A linear Curie-Weiss fit was done at high temperatures
(lines). The image in the inset shows a picture of a 1 x 1 x 0.15
mm$^{3}$ sized crystal with a mass of 1 mg.}}
\end{figure}

In Fig. \textcolor{black}{\ref{M-1}, we present the inverse susceptibility
of }Y$_{3}$Cu$_{9}$(OH)$_{19}$Cl$_{8}$\textcolor{black}{{} measured
on a single crystal for two different directions of the magnetic field.
In the temperature range 100-300~K, a clear Curie-Weiss behaviour
is observed with a large negative Weiss temperature of $\Theta_{W}=-100\,\text{K}$
for fields perpendicular to the kagome plane indicating a strongly
antiferromagnetic mean-exchange field. The effective moment for a
magnetic field applied perpendicular to the kagome plane is $\mu_{eff}\approx2.07\,\mu_{B}/\text{Cu}$
resulting in a Landé factor of $g\approx2.39$, assuming a simple
$J=S=1/2$ coupling system, typical for Cu$^{2+}$. For fields in
the kagome plane, the Curie-Weiss fit yields: $\mu_{eff}\approx1.89\,\mu_{B}/\text{Cu}$,
$\Theta_{W}=-86\,\text{K}$ and $g\approx2.18$. The leading exchange
coupling is most likely within the kagome plane, as the interplane-Cu-Cu
distances are much longer compared to the inplane distances. The Weiss
temperatures of }Y$_{3}$Cu$_{9}$(OH)$_{19}$Cl$_{8}$\textcolor{black}{{}
are quite similar to the results of W. Sun et al. for magnetic measurements
on powders of YCu$_{3}$(OH)$_{6}$Cl$_{3}$ with $\Theta_{W}=-99.2\,\text{K}$
\cite{Sun(2016)} indicating, that the leading exchange couplings
in the two Y-Cu-OH-Cl kagome systems are rather similar.}

\begin{figure}[h]
\includegraphics[width=1\columnwidth]{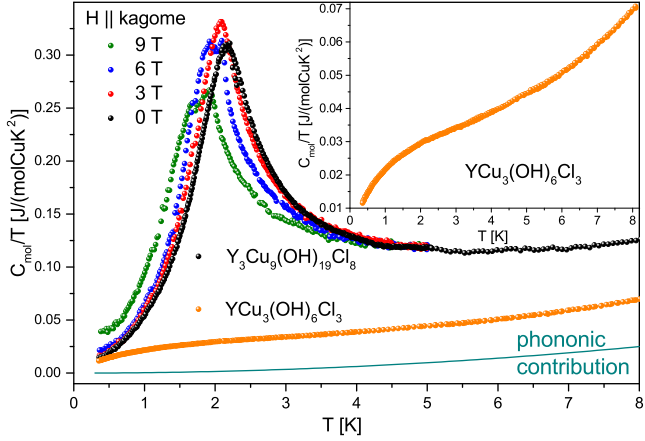}

\caption{\textcolor{black}{\footnotesize{}\label{HC}Specific-heat divided
by temperature at various magnetic fields of an impurity-free }{\footnotesize{}Y$_{3}$Cu$_{9}$(OH)$_{19}$Cl$_{8}$}\textcolor{black}{\footnotesize{}
single crystal and an YCu$_{3}$(OH)$_{6}$Cl$_{3}$ powder sample
with the P$\bar{3}$m1 structure. In the inset, the zero field curve
of YCu$_{3}$(OH)$_{6}$Cl$_{3}$ is enlarged and it is apparent,
that no long-range magnetic order occurs in the investigated temperature
range.}}
\end{figure}

\textcolor{black}{We performed specific-heat measurements on }Y$_{3}$Cu$_{9}$(OH)$_{19}$Cl$_{8}$\textcolor{black}{{}
single crystals in the temperature range from 0.35 to 270 K. While
all powder samples contain small amounts of Clinoatacamite only some
crystals have Clinoatacamite twinnings on their surface leading to
an anomaly at 6.5 K from the magnetic ordering of this impurity phase
\cite{Chu(2011),Norman(2016)}. The phase stability of }Y$_{3}$Cu$_{9}$(OH)$_{19}$Cl$_{8}$\textcolor{black}{{}
is close to that of Clinoatacamite and always both phases are formed
if the hydrothermal growth conditions are applied that are described
above. In magnetic measurements, the impurity contribution can even
dominate the signal and we have taken great care to select a crystal
without this impurity phase. In Fig. \ref{HC}, the temperature and
magnetic-field dependent specific heat of a single crystal without
any impurity phase is shown. In zero field (black curve) a maximum
is apparent at around 2.2 K. This maximum is slightly shifted to lower
temperatures with increasing field approaching $\sim2.0$~K at $\mu_{0}$H
$=9$~T, for H parallel to the kagome layer. A small shoulder appears
at the low temperature side for fields larger than 3~T. The entropy
gain within the maximum in the specific heat at zero field is $S=\stackrel[0.3]{4.5}{\intop}\frac{C_{mol}}{T}dT\approx0.59\frac{J}{mol\cdot K}\approx0.1R\cdot\text{ln}2$.
This shows that the ordered moment of }Y$_{3}$Cu$_{9}$(OH)$_{19}$Cl$_{8}$\textcolor{black}{{}
is strongly reduced and a large portion of the spin degrees of freedom
remain fluctuating. We have not subtracted any phononic contribution
for the entropy analysis, because at 4~K the estimated contribution
of the phonons amounts to only 5\%. The phononic contribution was
estimated in the range of 8~K to 26~K. The result of a linear fit
of $C_{mol}/T$ vs $T^{2}$ gives $\gamma_{0}=0.67(2)$~J/(molK$^{2}$)
and $\beta_{0}=3.38(6)$~mJ/(molK$^{4}$). This yields a Debye temperature
of $\theta_{D}\approx318\,\text{K}$. In addition, we do not observe
any phase transition in the temperature range from 8 to 270 K, in
agreement with the magnetic measurements and the temperature dependent
PXRD. The low ordering temperature together with the large Weiss-temperature
gives a frustration parameter $\Theta_{W}/T_{N}\approx50$, proving
that }Y$_{3}$Cu$_{9}$(OH)$_{19}$Cl$_{8}$\textcolor{black}{{} is
still a strongly frustrated material. Above T$_{N}$, the specific
heat of }Y$_{3}$Cu$_{9}$(OH)$_{19}$Cl$_{8}$\textcolor{black}{{}
follows a linear in T-dependence up to 8~K most likely due to the
enhanced spin fluctutations. }

\textcolor{black}{We have also measured the specific heat of a powder
sample of YCu$_{3}$(OH)$_{6}$Cl$_{3}$ (orange curves in Fig. \ref{HC}),
which clearly shows the absence of magnetic order down to 0.4~K which
is in agreement with the proposal of a spin-liquid ground state by
W. Sun et al. \cite{Sun(2016)}. Our low-temperature measurements
enhance the lower boundary of the frustration parameter of YCu$_{3}$(OH)$_{6}$Cl$_{3}$
to $\Theta_{W}/T_{N}>250$ and clearly proves the different magnetic
ground states of the two Y-Cu-OH-Cl compounds. The same linear fit
procedure was done for a specific heat measurement of YCu$_{3}$(OH)$_{6}$Cl$_{3}$
giving $\gamma_{0}=0.27(1)$~J/(molK$^{2}$), $\beta_{0}=1.22(3)$~mJ/(molK$^{4}$)
and $\theta_{D}\approx312\,\text{K}$, revealing the structural similarity
of these systems. Again, no phase transitions were observed up to
250~K. }

\begin{figure}[h]
\includegraphics[width=1\columnwidth]{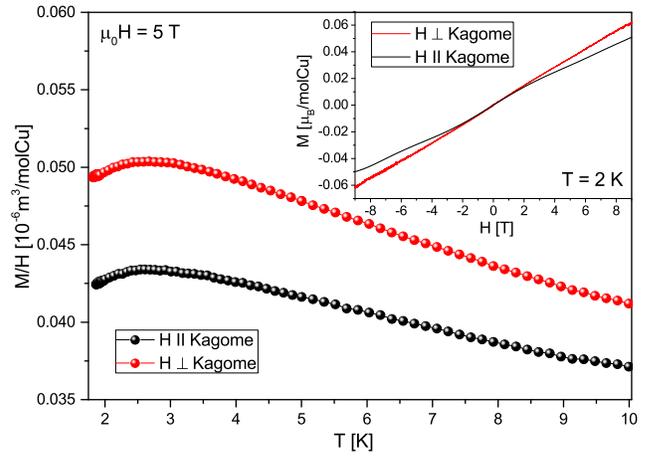}

\caption{{\footnotesize{}\label{M}Molar susceptibility of a Y$_{3}$Cu$_{9}$(OH)$_{19}$Cl$_{8}$
single crystal measured at 5 T from $T=10\,$K to 1.8~K. Inset: the
magnetization versus field curve measured at 2~K along and perpendicular
to the kagome layers.}}
\end{figure}

The magnetic transition at $T_{N}$ of Y$_{3}$Cu$_{9}$(OH)$_{19}$Cl$_{8}$
can also be determined by magnetic measurements in phase pure crystals,
which show a broad maximum at 2.5 K (see Fig. \ref{M}). This broad
transition is in contrast to the well-defined anomaly in the specific-heat
measurements and might be due to a large magnetic background because
of enhanced spin-fluctuations of that material. The magnetization
is larger for magnetic fields applied perpendicular to the kagome
planes, but the overall magnetic anisotropy is weak. The $M(H)$ curve
perpendicular to the field (inset of Fig. \ref{M}) shows a nearly
linear increase, while the parallel one has a small kink above 2~T.
This might be related to the field induced shoulder observed in the
specific heat data. Both $M(H)$ curves show no saturation up to 9~T.
In comparison with Herbertsmithite, the magnetic anisotropy in Y$_{3}$Cu$_{9}$(OH)$_{19}$Cl$_{8}$
is stronger but also favors the \textit{c}-direction as the easy magnetic
axis \cite{Han(2011)}.

\section{Conclusion}

In conclusion, Y$_{3}$Cu$_{9}$(OH)$_{19}$Cl$_{8}$ is a stoichiometric
quantum spin system with well-separated kagome layers of localized
Cu$^{2+}$ spins. Detailed structural refinements\textcolor{blue}{{}
}\textcolor{black}{of hydrothermally prepared single crystals} revealed
a crystal structure with two different copper positions, leading to
slightly anisotropic kagome layers. These structural modifications
were corroborated by optical absorption experiments which, in addition,
confirmed the chemical composition as the studied material shows insulating
behaviour with no indications of electron doping and the proposed
Dirac bands.

The partial release of magnetic frustration within the kagome layers
compared to YCu$_{3}$(OH)$_{6}$Cl$_{3}$ is also reflected in the
magnetic properties, because we observe weak but clear magnetic order
at $T_{N}=2.2$~K in magnetization and specific-heat measurements
on single crystals of Y$_{3}$Cu$_{9}$(OH)$_{19}$Cl$_{8}$. However,
the frustration effects are still very pronounced with a frustration
parameter of $\frac{\theta_{W}}{T_{N}}\sim50$. 

Low-temperature specific-heat measurements on a powder sample of YCu$_{3}$(OH)$_{6}$Cl$_{3}$
revealed the absence of magnetic order down to 0.4~K, leading to
a frustration parameter \textcolor{black}{$\Theta_{W}/T_{N}>250$}.
Therefore, the two Y-Cu-OH-Cl compounds present an unique setting
to investigate the change from a spin-liquid state to a strongly frustrated
AFM ordered state, by slightly releasing the frustration in a kagome
lattice via structural modification. Unlike for substitution series,
as e.g. Zn$_{x}$Cu$_{4-x}$(OH)$_{6}$Cl$_{2}$, where we always
encounter crystallographic disorder, the magnetic properties of the
two stochiometric compounds with fully occupied kagome sites might
be much more reliable with ab-initio calculations. Additionally, these
two stoichiometric kagome systems might open the way for a systematic
understanding of magnetic frustration in kagome materials, which would
require further more microscopic measurements of the spin-fluctuation
spectrum in these two systems. 
\begin{center}
\rule[0.5ex]{0.6\columnwidth}{0.5pt} 
\par\end{center}
\begin{acknowledgments}
The authors gratefully acknowledge support by the Deutsche Forschungsgemeinschaft
through grant SFB/TR 49. We thank Michael Seitz for the help with
the LA-ICP-MS measurements. This work is based on experiments performed
at the Swiss spallation neutron source SINQ, Paul Scherrer Institute,
Villigen, Switzerland.
\end{acknowledgments}

\end{document}